\documentclass[amsmath,amssymb,nofootinbib,twocolumn,preprintnumbers]{revtex4}
\pdfoutput=1

\usepackage{graphicx}
\usepackage{dcolumn}
\usepackage{hyperref}
\usepackage{bm}
\usepackage{amsmath}
\usepackage{amsfonts}
\usepackage{slashed}
\usepackage{hyperref}
\usepackage{color}

\def\ls{\mathrel{\lower4pt\vbox{\lineskip=0pt\baselineskip=0pt
           \hbox{$<$}\hbox{$\sim$}}}}
\def\gs{\mathrel{\lower4pt\vbox{\lineskip=0pt\baselineskip=0pt
           \hbox{$>$}\hbox{$\sim$}}}}
\def\drawbox#1#2{\hrule height#2pt
\hbox{\vrule width#2pt height#1pt \kern#1pt
              \vrule width#2pt}
              \hrule height#2pt}

\def\Asym#1#2{\vcenter{\vbox{\drawbox{#1}{#2}
              \kern-#2pt       
              \drawbox{#1}{#2}}}}

\def\nn{\nonumber}

\newcommand{\be}{\begin{equation}}
\newcommand{\ee}{\end{equation}}
\newcommand{\bea}{\begin{eqnarray}}
\newcommand{\eea}{\end{eqnarray}}

\newcommand{\gsim}{\lower.7ex\hbox{$\;\stackrel{\textstyle>}{\sim}\;$}}
\newcommand{\lsim}{\lower.7ex\hbox{$\;\stackrel{\textstyle<}{\sim}\;$}}

\newcommand{\ben}{\begin{enumerate}}
\newcommand{\een}{\end{enumerate}}
\newcommand{\bei}{\begin{itemize}}
\newcommand{\eei}{\end{itemize}}

\begin{document}

\preprint{MI-TH-1617}
\title{Exploring the Jet Multiplicity in the 750 GeV Diphoton Excess}

\author{Mykhailo Dalchenko$^{1}$}
\author{Bhaskar Dutta$^{1}$}
\author{Yu Gao$^{1}$}
\author{Tathagata Ghosh$^{1}$}
\author{Teruki Kamon$^{1,2}$}

\affiliation{
$^1$Mitchell Institute for Fundamental Physics and Astronomy,
Department of Physics and Astronomy, Texas A\&M University, College Station, TX 77843-4242, USA}
\affiliation{$^{2}$~Department of Physics, Kyungpook National University, Daegu 702-701, South Korea}

\begin{abstract}
The recent diphoton excess at the LHC has been explained tentatively by a Standard Model (SM) singlet scalar of 750 GeV in mass, in  the association of heavy particles with SM gauge charges. These new particles with various SM gauge charges induce loop-level couplings of the new scalar to $WW$, $ZZ$, $Z\gamma$, $\gamma\gamma$, and $gg$. We show that the strength of the couplings to the gauge bosons also determines the production mechanism of the scalar particle via $WW,\, ZZ,\, Z\gamma,\, \gamma\gamma,\, gg$ fusion which leads to individually distinguishable jet distributions in the final state where the statistics will be improved in the ongoing run. The number of jets and the leading jet's transverse momentum distribution in the excess region of the diphoton signal can be used to determine the coupling of the scalar to the gauge bosons arising from the protons which subsequently determine the charges of the heavy particles that arise from various well-motivated models.  
\end{abstract}

\maketitle

\label{intro}

Both ATLAS and CMS collaborations have reported
an excess of diphoton events at a reconstructed invariant mass of about 750 GeV.
This excess is visible in the data at 13 TeV~\cite{ATLAS_diphoton:2015, CMS_diphoton:2015}
and consistent with 8 TeV~\cite{bib:CMS_Moriond, bib:ATLAS_Moriond}.
The local signal significance of the excess by ATLAS is $3.6\sigma$ for an integrated luminosity of 3.2 ${\rm fb}^{-1}$ at 13 ~TeV 
and about 1.9 $\sigma$ from 20.3 ${\rm fb}^{-1}$ at 8 TeV, while the local signal significance by CMS is
3.4$\sigma$ by combining results from  luminosities of 3.3 ${\rm fb}^{-1}$ and 19.7 ${\rm fb}^{-1}$ at 13 and 8 TeV, respectively.
It is noted that the observed significance by CMS is maximized for a narrow decay width of $\Gamma/m\leq 10^{-2}$, while the ATLAS result
is in favor of a larger width with $\Gamma/m\sim 0.06$.   
Using the limited data, ATLAS  has also reported jet multiplicity distributions in the diphoton excess region and its sidebands.  

In this Letter, we point out that the jet topology could be powerful in distinguishing different models in the excess region when more data becomes available in the ongoing run.

\begin{figure}[h]
\includegraphics[scale=0.4]{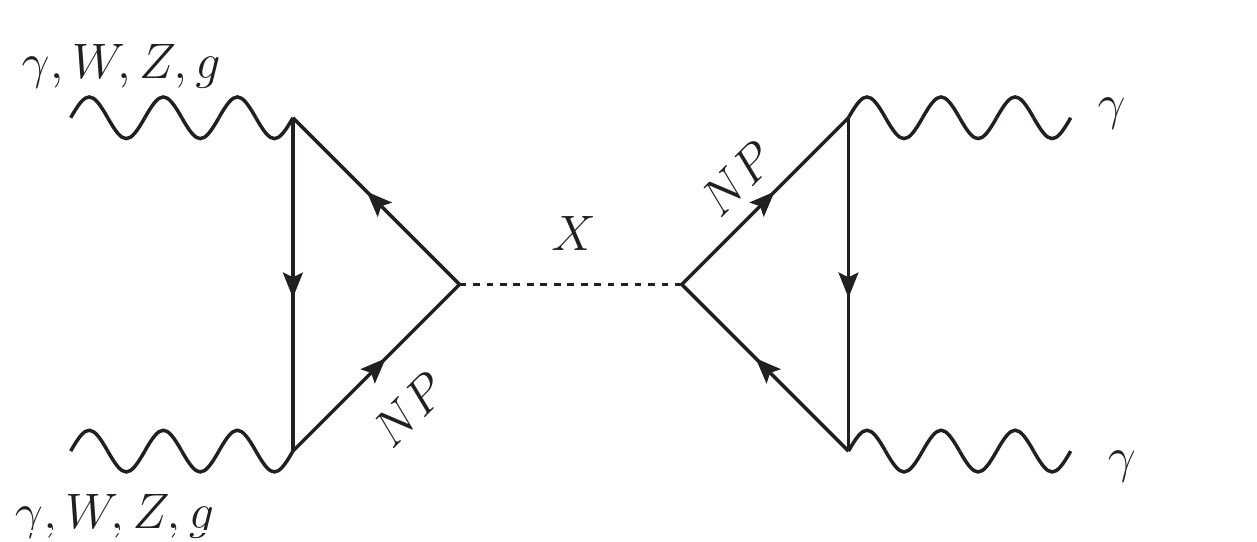}
\caption{Loop-induced couplings between photons and 750 GeV resonance. `NP' denotes for any new physics heavy particles that are charged under the SM gauge groups.}
\label{fig:feynman}
\end{figure} 
 
Among numerous hypotheses~\cite{bib:early, Nakai:2015ptz} to explain the diphoton excess, we focus on the phenomenologically minimal setup by introducing a Standard Model(SM) singlet $X$ with a mass of 750~GeV accompanied by multiplets of vector-like particles which possess SM charges. 
The effective coupling of SM SU(2)$_L$ and U(1)$_Y$ gauge bosons to the 750 GeV resonance can be induced at loop level by heavy new particles, as shown in Fig.~\ref{fig:feynman}, and can be written as
\be 
{\cal L }_{\text{eff}}\supset
\kappa_1 X B^{\mu\nu}B_{\mu\nu}+\kappa_2 X W^{\mu\nu}W_{\mu\nu}+\kappa_3 X G^{\mu\nu}G_{\mu\nu}.
\ee

The coupling values are determined from the heavy particles' masses and charges under the SM gauge groups. After rotations they give rise to effective couplings between $X$ and the physical gauge bosons,
\bea
 \kappa_{\gamma \gamma } & = & \kappa_1 \cos^{2} \theta_W + \kappa_2 \sin^{2} \theta_W \, , \nn \\
 \kappa_{Z Z } & = &  \kappa_2 \cos^{2} \theta_W + \kappa_1 \sin^{2} \theta_W \, , \\
 \kappa_{Z \gamma } & = &  (\kappa_2 - \kappa_1) \sin 2 \theta_W \, , \nn \\
 \kappa_{WW } & = & \kappa_2 \, ,\ \ ~ \kappa_{gg }  = \kappa_3 \nn
\label{eq:eff_couplings_singlets}
\eea

The relative size of these couplings are among the most characteristic predictions of new physics scenarios that implement new heavy particles. For instance, Ref.~\cite{Dutta:2016ach} proposes the gauge unification under SU(6) at ${\cal O}(10^{16})$ GeV which require the existence of  a number of new fermions which include down type SU(2) singlet vector-like quark $D$ and vector-like SU(2) lepton doublet $L$.  The multiplicity of these new fermions and masses fixes $\kappa_i$ and subsequent $X$ decay branching fractions into $\gamma\gamma,gg,Z\gamma,ZZ$ and $WW$ final states. Similarly, one can introduce $Q$, $E$ and $U$ type vector-like fermions to satisfy the data in the context of $10+\overline{10}$ representation of SU(5)~\cite{bib:SU5}. These are just two examples of new physics models in which the heavy particles' SM charge assignments predict the relative sizes of $\kappa_i$.

The relative strengths of $\kappa_i$s not only predict the branching ratios of $\gamma\gamma,\,gg,\,Z\gamma,\,ZZ$ and $WW$ final states, but they also give us several production possibilities of $X$ via fusion of different gauge bosons. For example, in case of larger $\kappa_{WW,\,ZZ}$ couplings, we expect $pp\rightarrow ZZ,Z\gamma,WW,jj$ with high $p_T$ jets to be  the primary predictions from these effective couplings. Such channels can be tested in the upcoming LHC dijet, multi-lepton, and leptons+photon resonance searches. Alternatively, in a dominant $\kappa_{\gamma\gamma}$ case we can expect less associated jets with significant $p_T$.
Therefore, different scenarios with various values of $\kappa_i$ that yield unique final state jet distributions provide us with a very promising probe of the production mechanism of the resonance.

Let us now discuss the productions of $X$ via various mechanisms due to each $\kappa_i$ and their predictions on the associated jet multiplicity and the leading jet $p_T$ distribution.

\label{aa}
~{\bf Photon-fusion} has been proposed in Ref.~\cite{Csaki:2016raa, He:2016olo}, and the initial state photon is studied kinematically in detail~\cite{Harland-Lang:2016qjy}. These studies differentiate the photon fusion from other, especially gluon fusion, in the photon kinematics and jet $p_T$. In this work, we focus on the jet multiplicity as the distinctive feature and compare with experimental data. 

Photon-fusion can obtain domination with $\kappa_1\gg \kappa_2,\kappa_3$, for example, when the mediators inside the loop are non-colored SU(2)$_L$ singlets, like heavy partners of right-handed charged leptons. Their electric charge (from their hypercharge) generates photonic couplings that explain the excess, yet without inciting large couplings to the gluon or $W$ boson.

\begin{figure}[h]
\includegraphics[height=4cm]{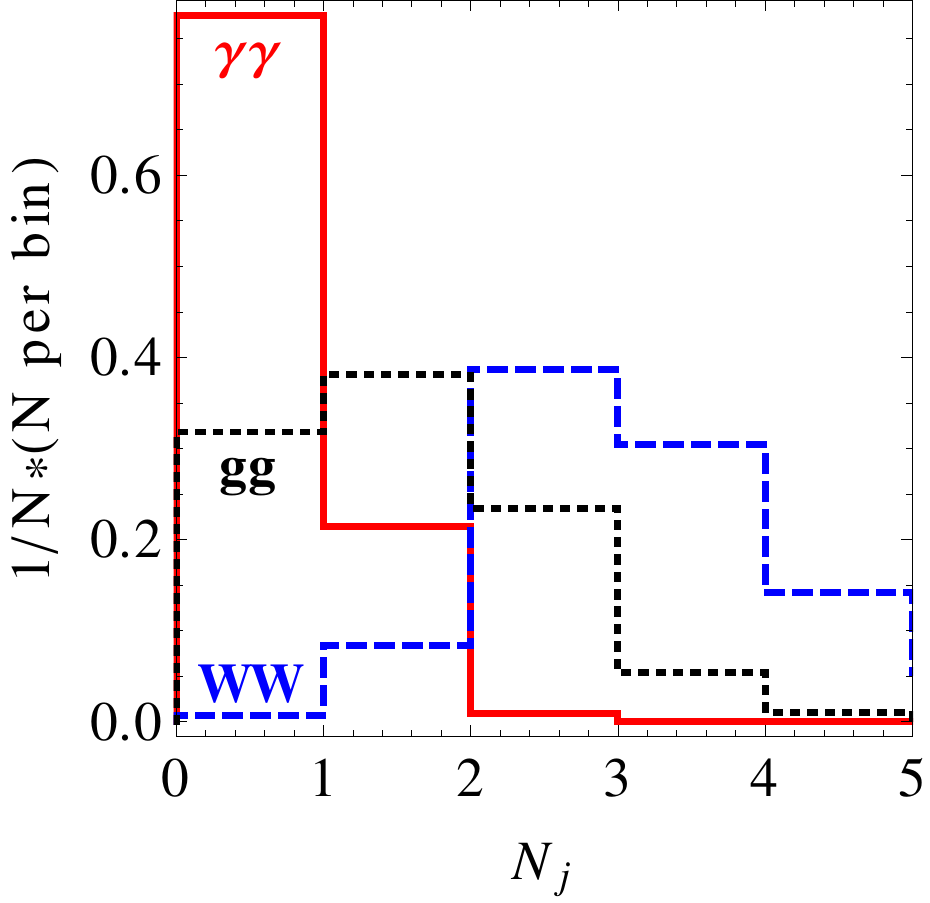}
\includegraphics[height=3.9cm]{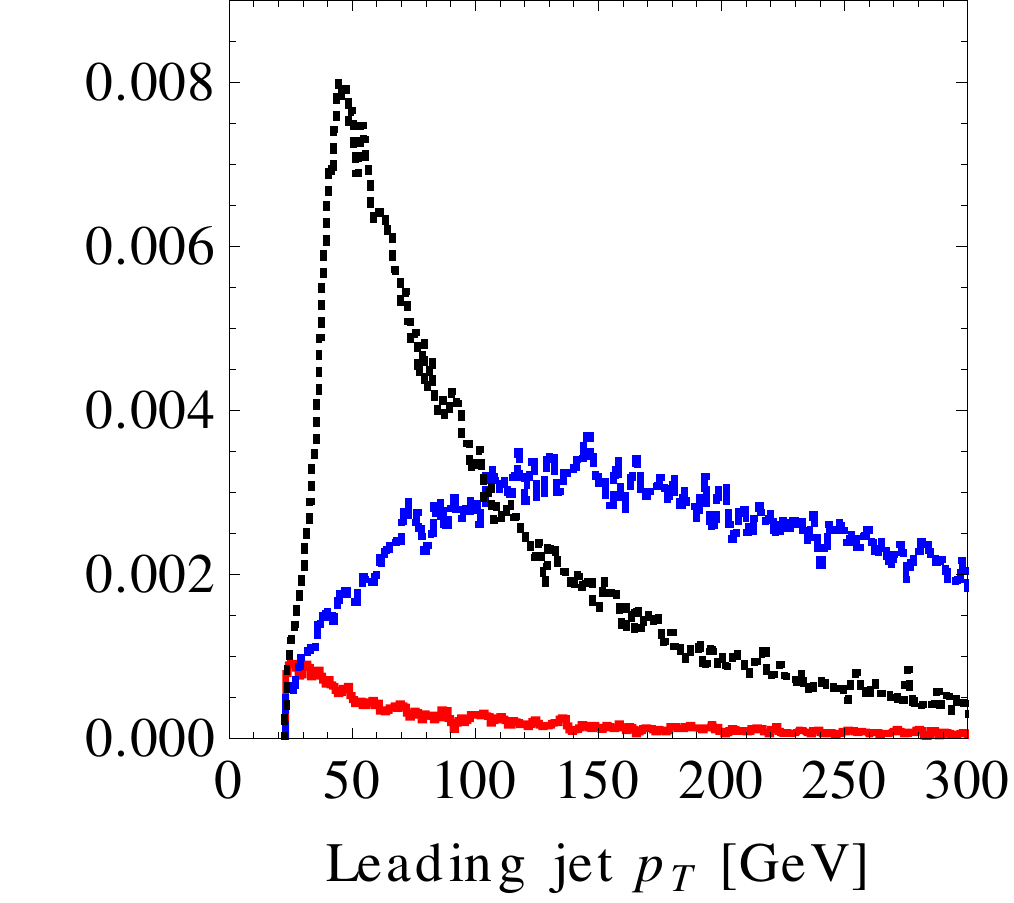}
\caption{Detector-level central region $N_j$(left) and jet $p_T$ distribution from photon (red), $WW$ (blue-dashed) and $gg$ (black-dotted) fusions.}
\label{fig:Nj_pt}
\end{figure}

A unique kinematic feature of inelastic photon fusion is the photon's collinear enhancement that strongly favours low $p_T$ recoil on its parent parton, thus leads to low $p_T$ and/or high pseudorapidity initial state jets, which prevent such jets from populating the central detector regions. Elastic photons~\cite{Csaki:2016raa,Harland-Lang:2016qjy} from the proton also make a sub-leading contribution, where photon emission is also suppressed by proton scattering $p_T$ and would dominantly yield a jet-less final state. Comparing the number of jets, $N_j$ in the observed $\eta$ ranges from ATLAS's diphoton sample~\cite{bib:ATLAS_Moriond} can be a useful way to check whether the experimental excesses show preference to photon-fusion as the major production mechanism.

We simulated the $N_j$ distribution with a $\kappa_1$ dominant benchmark point for the jet distributions from photon fusion. We use MadGraph~\cite{bib:madgraph5} and the recent NNPDF23\_LO set~\cite{bib:nnpdf}, which includes the photon's distribution inside the proton. Pythia~\cite{bib:pythia} is used for parton shower and Delphes~\cite{bib:delphes} for detector simulation. The $N_j$ from photon-fusion is shown in Fig~\ref{fig:Nj_pt}, with jet $p_T>25$~GeV in the pseudorapidity range $|\eta|<4.4$, from ATLAS~\cite{bib:ATLAS_Moriond}. We consider the pile-up effects to be well-eliminated in experimental analysis and are thus not included here. To fully account for the jets from $\gamma\gamma$ fusion, we calculate both the total cross section $\sigma^{\text{tot}}(\gamma\gamma\rightarrow\gamma\gamma)$ where the initial $\gamma$ as a parton, and also for the one-jet process $\sigma(\gamma p\rightarrow\gamma\gamma j)$, plus subsequent showers. The latter gives the $N_j$ distribution of one or more jets. We then add the difference $\Delta \sigma\equiv \sigma^{\text{tot}}-\sigma^{ap}$, which fails to produce a jet, to the zero-jet bin. We did not include the elastic and semi-elastic contributions here, which would also dominantly fall into the zero-jet case.  

As a note, $\kappa_1$ would also enable $ZZ$ fusion, and $\kappa_{ZZ}/\kappa_{\gamma\gamma}$ could be enhanced in the limit $\kappa_1 \rightarrow -\kappa_2 \tan^2\theta_W$, however, this condition requires a tuning of gauge group mixing and may be very model-dependent. Generally the collinear enhancement of $\gamma$ emission would let $\gamma\gamma$ fusion dominate over $ZZ$ fusion (and the mixed $\gamma Z$). As $ZZ$ is kinematically almost identical to $WW$ fusion, we do not list it as a separate case in this work. Both $Z\gamma$ and $Z Z$ can lead to  high $p_T$ jet(s).

\label{WW}
~{\bf $\bf{WW}$ fusion}, in comparison, is present if the heavy mediators are charged under the SM's SU(2)$_L$, e.g.,  vector-like lepton doublets, quark doublets in $5+\bar{5}$ and $10 +\bar{10}$ multiplets of SU(5), etc.

Noted that with a non-zero $\kappa_2$, $\kappa_{WW}$ would often coexist with $\kappa_{\gamma\gamma}$, and $WW, \gamma\gamma$ fusions would interfere. For illustrative purposes, here we choose a special case $\kappa_1\sim -\kappa_2 \tan^2\theta_W$ to suppress $\kappa_{\gamma\gamma}$ relative to $\kappa_{WW}$, and provide a $WW$ fusion dominated production process.

Unlike the $\gamma\gamma$ fusion case, the $WW$ fusion always comes with two associated initial state jets (aka VBF jets) and the central jet multiplicity would peak at $N_j=2$. Due to the weak-scale mass of the $W$ boson, $W$s are not forwardly enchanced, and a typical ISR jet would acquire $p_T\sim {\cal O}(M_W)$ or higher, as shown in Fig.~\ref{fig:Nj_pt}. It is clear that $WW$ and $\gamma\gamma$ fusion cases differ significantly in both jet multiplicity and jet $p_T$ distributions.

\label{gg}
~$\bf{gg}$~{\bf fusion} can be the leading production channel if heavy mediator are colored, or if $X$ is a composite particle made of colored fields~\cite{Nakai:2015ptz, Bai:2015nbs}, etc. A very similar case is that $X$ may have a small tree-level coupling $X\bar{q}q$ with quarks, if it is an SU(2)$_L$ doublet. Both $gg, q\bar{q} \rightarrow X$ initial states are dominated by QCD and produce ample initial state radiation (ISR) jets. The jet multiplicity distribution will follow a power-law shape which is typical for QCD radiation, and  jet $p_T$ distribution will also show a similar behavior, as shown in Fig.~\ref{fig:Nj_pt}. 

\begin{figure}[h]
\includegraphics[height=3.2cm]{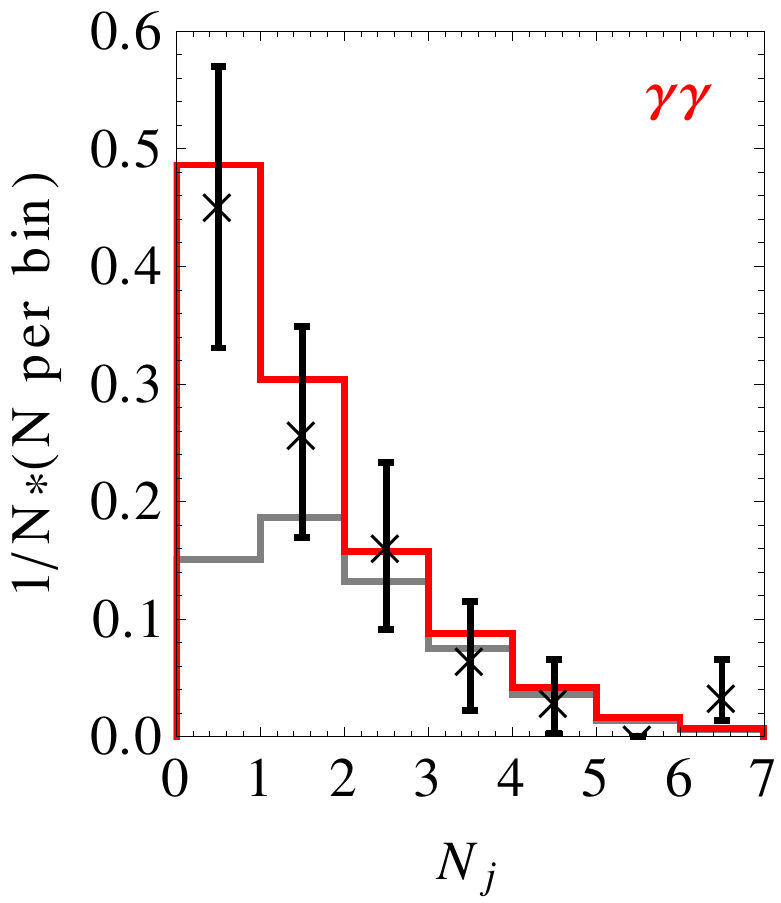}
\includegraphics[height=3.2cm]{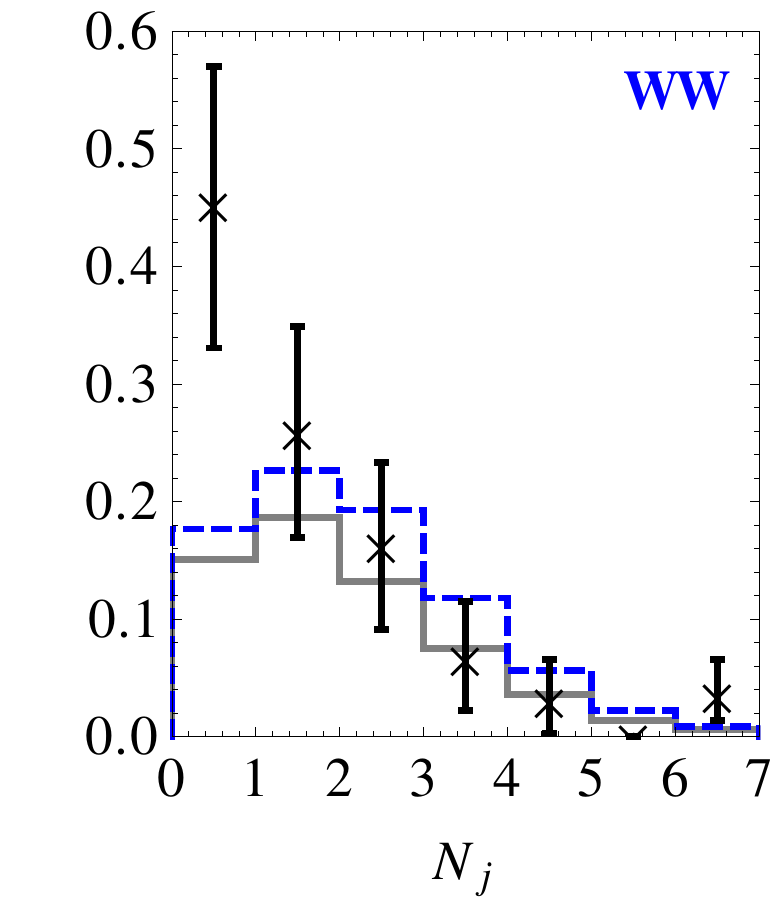}
\includegraphics[height=3.2cm]{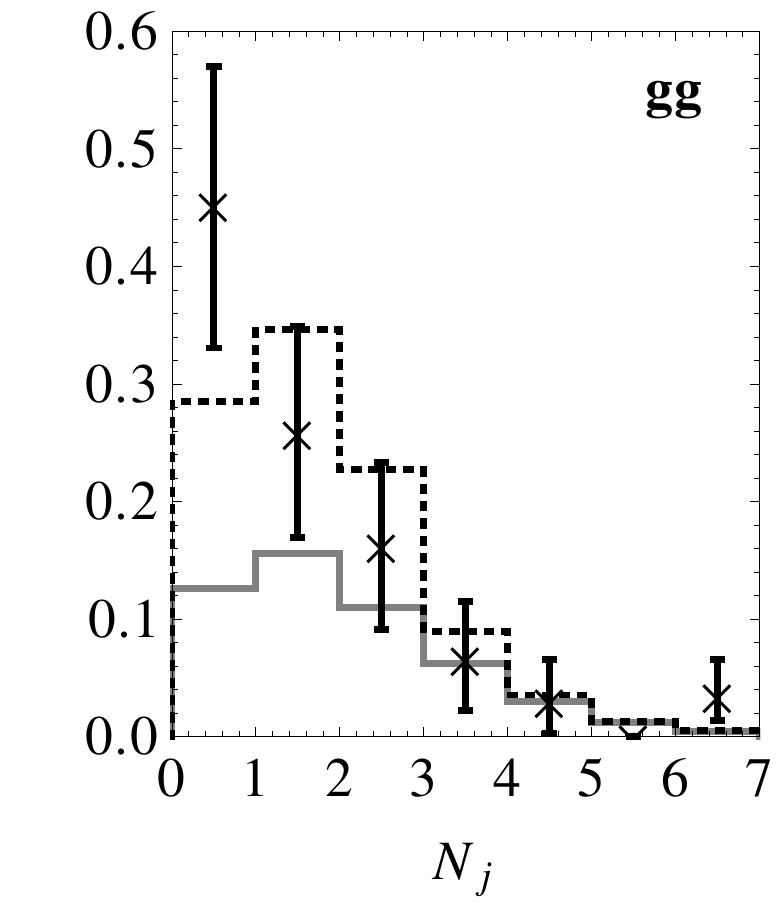}
\caption{Signal in $N_j$ distribution for $\gamma\gamma, WW$ and $gg$ fusion channels with ATLAS's diphoton data~\cite{bib:ATLAS_Moriond} Spin-0 criteria are assumed. The shape of re-scaled SM background (gray) is taken from the 700-840 GeV $M_{\gamma\gamma}$ Monte Carlo within ATLAS's report.}
\label{fig:fit2ALTAS_Nj}
\end{figure}

In the event generation for $gg$ fusion, we included $gg\rightarrow X, gp\rightarrow Xj$ and  $pp\rightarrow Xjj$ processes to fully account for ISR jets. MLM matching~\cite{Mangano:2006rw} with  $xqcut > 40$ GeV and $Qcut > 40$ GeV are used to avoid double counting. Although both $N_j$ distributions favor low multiplicity in both $\gamma\gamma, gg$ fusion, $gg$ differs from $\gamma\gamma$ with a much less pronounced weight in the zero-jet final state. This indicates that $gg$ fusion production has a higher fraction of signal diphoton events with ISR jet(s), whereas $\gamma\gamma$ fusion predicts much fewer associated jets in signal events. Fusion of massive gauge bosons, $WW$ and $ZZ$, also predicts associated jets with $p_T$ above their mass scale, and a different shape in the leading jet $p_T$ spectrum, as shown in Fig.~\ref{fig:Nj_pt}.

\label{fit to data}
{\bf Experimental ${\bf N_j}$ data} from the recent ATLAS report~\cite{bib:ATLAS_Moriond} shows that the excess of events under the selection rule of a spin-0 resonance in the 700-840 GeV mass range, mostly fall into the zero-jet bin. In order to make a fit to the $N_j$ data, in Fig.~\ref{fig:fit2ALTAS_Nj} we allow the signal-to-background ratio S/B to float for different channels, which indicates the size of an excess versus a fixed-rate SM background.  It is clear that the photon fusion mechanism makes the best match to the $N_j$ shape due to its lack of central jets. A value of  S/B = 0.6 makes a best-fit for $\gamma\gamma$ fusion, and S/B $\sim 1$ for $gg$ fusion which gives a worse fit in comparison. Interestingly, a background-only shape is strongly favored in the fit of $WW$ fusion, due to the higher $N_j$ in $WW$ fusion.

\begin{figure}[h]
\includegraphics[scale=0.56]{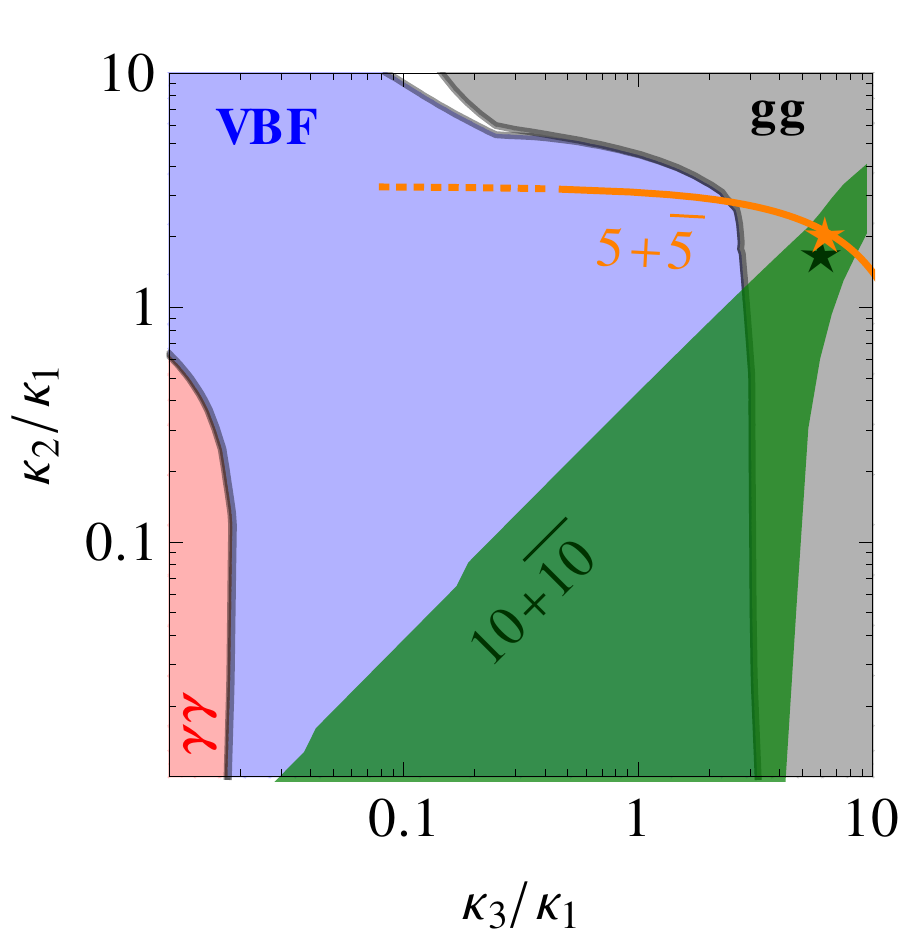}
\caption{Regions where $\gamma\gamma, WW/ZZ, gg$ fusion individually become the dominant production mechanism. 
Sample $5+\overline{5}$ (orange), $10+\overline{10}$ (green) parameter regions are also shown for various $r_{L/D}$, and the dashed ends can extend to the large $r_{L/D}\gg 1$ range. The asterisk symbols mark the high scale unification point where $r_{L/D}=0.95$ using $5+\overline{5}$.  The $10+\overline{10}$ region is extended due to the difference between $r_{E/U}$ and $r_{E/Q}$. In both scenarios, all heavy vector fermions assume their masses where unification is achieved. For $WW/ZZ$ fusion we use the $VBF$ cross section with a leading jet $p_T>50$GeV cut. 
}
\label{fig:photon_domination}
\end{figure}

If the diphoton excess is explained by photon fusion, the direct prediction on jets, in case they exist, will be of low $p_T$ due to the forward enhancement of photon emissions. A photon fusion dominated production would also suggest enhanced $X$ decay branching fraction into $\gamma\gamma$ (also yet not necessarily $Z\gamma, ZZ$) final states, while $X\rightarrow WW,jj$ would be suppressed. Similarly, if $WW$, $gg$ channels were to dominate the production, the corresponding $X$ decays would also become dominant and could be tested with upcoming LHC data. It is worthwhile to note that the  $\gamma\gamma$ fusion cross section scales with the beam energy differently than from that of $gg$ fusion, i.e., by a factor $\sim 0.4$.  We find that $\sigma_{\gamma \gamma}$ at 8 TeV is $\sim 1.9$ times smaller than the same  at 13 TeV which is comparable to $\sigma_{q\bar{q}}$. In contrast, the $\sigma_{g g}$ at 8 TeV is $\sim 4.7$ times smaller than the same at 13 TeV. Consequently the production cross section for $\gamma\gamma$ initial states is compatible with more than 2$\sigma$ deviation for the ATLAS and less than 2$\sigma$ deviation for the CMS 8 TeV results. 
However, we require more data to understand the consistency between the 8 and 13 TeV data conclusively.

In Fig~\ref{fig:photon_domination}, we show the parameter regions where each of $\gamma\gamma, WW/ZZ, gg$ fusion individually become the dominant (more than $50\%$ of the sum of all three cross sections) production channel. Here we consider the total cross section of $\gamma\gamma\rightarrow X$ for photon fusion, the VBF cross section $pp\rightarrow Xjj$ with a leading jet $p_T>50$ GeV cut that selects $W/Z$ fusion, and $gg\rightarrow X$ for gluon fusion. Note that the QCD-dominated $gg$ fusion benefits from inclusion of ISR jets that open more initial states, thus the realistic $gg$ cross section would scale up by a ${\cal O}(1)$ factor and could cause the $gg$/VBF boundary to move slightly left. 

It is interesting to note the difference in $\kappa_i$ between various new physics scenarios. The `reduced' parameter range of effective couplings with \{D, L\} fermions in $5+\overline{5}$ and \{Q, U, E\} in $10+\overline{10}$ scenarios in the context of SU(5) is  shown in Fig~\ref{fig:photon_domination}. The vector-like fermion contributions to the effective couplings are proportional to $\lambda_f/M_f$, where $\lambda_f$ denotes their coupling to $X$ via $\lambda_f X \bar{f} f$. In the `reduced' parameter space, we neglect the sub-dominant dependence of fermion mass in the loop factors and consider a lepton-quark ratio $r_{L/D}\equiv \frac{\lambda_L}{M_L}/\frac{\lambda_D}{M_D}$. Using $5+\overline{5}$ fields, after choosing a value of $r_{L/D}$, we find that the dependence on fermion mass becomes less sensitive, and the relevant parameter space can be approximated by a curve (orange). The high scale grand unification scenario  occurs at 
$r_{L/D}=0.95$, where we use renormalization group equations to determine the values of $\lambda$ and $M$ at the TeV scale, starting from the grand unified theory (GUT) scale, showing this unification with an asterisk symbol. For illustrative purposes, we assume common fermion mass at GUT scale and fix low-scale fermion masses at $M_L=400$ GeV and $M_D=766$ GeV. In the $10+\overline{10}$ case, the parameter space broadens due to the fact that $Q, U$ fields can assume different $\lambda/M$. Even away from the unification point, one can use the component fields of $5+\overline{5}$, $10+\overline{10}$ to explain the excess. Due to the presence of vector-like quarks, the gluon initial state dominates in the gauge unification model. However, if we use the $E+\bar{E}$ fields to be light then $\kappa_1$ can dominate.  

The discussions so far have not taken into account the gauge boson fusion's impact on the size of $X$ decay width. While CMS~\cite{bib:CMS_Moriond} reported a slight favor for a narrow $X$ width, both CMS and ATLAS results are consistent with a large width scenario up to a few percent of the $X$ mass. 


A promising way to increase the $X$ width is to couple $X$ with complete SM singlets, e.g. via a $X\bar{N}N$ type interaction. Particle $N$ avoids the detection at the LHC and makes a significant contribution to $\Gamma_X$ as invisible width. Such a scenario  also faces problems from monojet searches~\cite{bib:monojet_experiment}. The monojet channel is enhanced by BF$(X\rightarrow\bar{N}N)$/BF$(X\rightarrow \gamma\gamma)$ which can be quite significant if 
$\bar{N}{N}$ is the major contributor to a large $X$ width. 

For the photon fusion, only about {2-3\%} of $\gamma\gamma\rightarrow X$ provide a leading jet more than 100 GeV and missing transverse energy of {200} GeV to be registered as a monojet event. For $WW$ and $gg$ fusion cases, this fraction is {27}\% and {4}\%, respectively. Therefore, $\gamma\gamma$ fusion is slightly better than $gg$ in terms of monojet bound, while $WW$ can be significantly worse.

\begin{table}
\begin{tabular}{c|c|c|c}
\hline
Production channel  &\ $\gamma\gamma$\ &\ $WW/ZZ$\ &\ $gg/q\bar{q}$\ \\
\hline
Probable $N_j$ ($|\eta| < 4.4$) & $\sim 0$ & $2-3$ & $1-2$ \\
Leading jet $p_T$ & $\lsim 10$ GeV & $\gtrsim M_{W}$ & QCD-like \\
Monojet constraint$^{*}$ & Yes & Severe & Yes \\ 
\hline
\end{tabular}
\caption{Characteristics of jets in different vector gauge boson fusion (VBF) production mechanisms. The non-VBF $q\bar{q}$ process is listed under VBF $gg$ process due to similarity in their kinematics.\\ $^*$ \ if a large invisible width is present.}
\label{tab:summary}
\end{table}

{\bf To summarize,} we have investigated  the possibilities of distinguishing different models based on the number and $p_T$  of jets in the excess regions which can be confirmed when more data is available in the ongoing run.  The 750 GeV resonance has been explained using a SM singlet and vector-like particles with different SM charges in various well-motivated models. Based on the SM charge assignments, the production mechanism of these particles at the LHC can occur via $\gamma\gamma$, $WW/ZZ$, $gg$ initial states. The jet spectrum associated with these different production processes can be different, which we summarize  in Table~\ref{tab:summary}. We also found that the photon fusion initial state matches the $N_j$ shape  provided by ATLAS well, due to its lack of central jets whereas the $gg$ fusion provides a worse fit in comparison. The $WW$ fusion initial state fit is lot worse due to the central jet multiplicity peaking at 2. With more data, we find that the associated jet spectrum will be able to distinguish different models which explain the diphoton excess.


{\bf Acknowledgements.}
The authors thank Doojin Kim and Joel Walker for helpful discussions. B.D. and T.K. are partially supported by DOE Grant DE-FG02-13ER42020. M.D. and Y.G. thank the Mitchell Institute for Fundamental Physics and Astronomy for support. T.K. is also supported in part by Qatar National Research Fund under project NPRP 5-464-1-080.


\begin{thebibliography}{99}

\bibitem{ATLAS_diphoton:2015}
  ATLAS collaboration,
  ATLAS-CONF-2015-081.
  
\bibitem{CMS_diphoton:2015}
  CMS Collaboration,
  CMS-PAS-EXO-15-004.

\bibitem{bib:CMS_Moriond}
  CMS Collaboration,
  CMS-PAS-EXO-16-018.

\bibitem{bib:ATLAS_Moriond}
  ATLAS collaboration,
  ATLAS-CONF-2016-018.


\bibitem{bib:early}
  K.~Harigaya and Y.~Nomura,
  Phys.\ Lett.\ B {\bf 754}, 151 (2016);
  M.~Backovic, A.~Mariotti and D.~Redigolo,
  JHEP {\bf 1603}, 157 (2016);
  A.~Angelescu, A.~Djouadi and G.~Moreau,
  Phys.\ Lett.\ B {\bf 756}, 126 (2016);
S.~Knapen, T.~Melia, M.~Papucci and K.~Zurek,
 Phys.\ Rev.\ D {\bf 93}, no. 7, 075020 (2016)
  D.~Buttazzo, A.~Greljo and D.~Marzocca,
  Eur.\ Phys.\ J.\ C {\bf 76}, no. 3, 116 (2016);
  A.~Pilaftsis,
  Phys.\ Rev.\ D {\bf 93}, no. 1, 015017 (2016);
  S.~D.~McDermott, P.~Meade and H.~Ramani,
  Phys.\ Lett.\ B {\bf 755}, 353 (2016);
  A.~Kobakhidze, F.~Wang, L.~Wu, J.~M.~Yang and M.~Zhang,
  Phys.\ Lett.\ B {\bf 757}, 92 (2016);
R.~Franceschini {\it et al.},
JHEP {\bf 1603}, 144 (2016);
  S.~Fichet, G.~von Gersdorff and C.~Royon,
  Phys.\ Rev.\ D {\bf 93}, 075031 (2016);
  B.~Dutta, Y.~Gao, T.~Ghosh, I.~Gogoladze and T.~Li,
  Phys.\ Rev.\ D {\bf 93}, no. 5, 055032 (2016);
  P.~S.~B.~Dev, R.~N.~Mohapatra and Y.~Zhang,
  JHEP {\bf 1602}, 186 (2016)
  G.~M.~Pelaggi, A.~Strumia and E.~Vigiani,
  JHEP {\bf 1603}, 025 (2016)

\bibitem{Nakai:2015ptz}
  Y.~Nakai, R.~Sato and K.~Tobioka,
   Phys.\ Rev.\ Lett.\  {\bf 116}, no. 15, 151802 (2016);

\bibitem{Dutta:2016ach} 
  B.~Dutta, Y.~Gao, T.~Ghosh, I.~Gogoladze, T.~Li and J.~W.~Walker,
  arXiv:1604.07838 [hep-ph].
  
\bibitem{bib:SU5} 
  L.~J.~Hall, K.~Harigaya and Y.~Nomura,
  JHEP {\bf 1603}, 017 (2016)
  B.~Dutta, Y.~Gao, T.~Ghosh, I.~Gogoladze, T.~Li, Q.~Shafi and J.~W.~Walker,
  arXiv:1601.00866 [hep-ph].
  


\bibitem{Csaki:2016raa} 
  C.~Csaki, J.~Hubisz, S.~Lombardo and J.~Terning,
  Phys.\ Rev.\ D {\bf 93}, 095020 (2016)
  doi:10.1103/PhysRevD.93.095020
\bibitem{He:2016olo} 
  M.~He, X.~G.~He and Y.~Tang,
Phys. Lett. B  {\bf 759}, 166 (2016),
  doi:10.1016/j.physletb.2016.05.056





\bibitem{Harland-Lang:2016qjy} 
  L.~A.~Harland-Lang, V.~A.~Khoze and M.~G.~Ryskin,
  JHEP {\bf 1603}, 182 (2016)
  doi:10.1007/JHEP03(2016)182


\bibitem{bib:madgraph5}
J.~Alwall, 
  {\it et al.},
``Madgraph 5: going beyond'',
JHEP {\bf 06}, 128 (2011);
  J.~Alwall,
  {\it et al.},
  JHEP {\bf 07}, 079 (2014)

\bibitem{bib:nnpdf}
  R.~D.~Ball {\it et al.},
  Nucl.\ Phys.\ B {\bf 867}, 244 (2013)
  doi:10.1016/j.nuclphysb.2012.10.003

\bibitem{bib:pythia}
T.~Sjöstrand, 
{\it et al.},
  Comput.\ Phys.\ Commun.\  {\bf 191}, 159 (2015)

\bibitem{bib:delphes}
    J. de Favereau {\it et al.} [DELPHES 3 Collaboration], 
JHEP {\bf 02}, 057 (2014).

\bibitem{Bai:2015nbs} 
  Y.~Bai, J.~Berger and R.~Lu,
  Phys.\ Rev.\ D {\bf 93}, no. 7, 076009 (2016)
  doi:10.1103/PhysRevD.93.076009
  J.~Cao, C.~Han, L.~Shang, W.~Su, J.~M.~Yang and Y.~Zhang,
  Phys.\ Lett.\ B {\bf 755}, 456 (2016)
  doi:10.1016/j.physletb.2016.02.045
  R.~Ding, Z.~L.~Han, Y.~Liao and X.~D.~Ma,
  Eur.\ Phys.\ J.\ C {\bf 76}, no. 4, 204 (2016)
  doi:10.1140/epjc/s10052-016-4052-6

\bibitem{Mangano:2006rw} 
  M.~L.~Mangano, M.~Moretti, F.~Piccinini and M.~Treccani,
  JHEP {\bf  01} (2007) 013
  [hep-ph/0611129].


\bibitem{bib:monojet_experiment}
CMS Collaboration, CMS-PAS-EXO-15-003.


\end{thebibliography}
\end{document}